\documentclass[12pt]{article}
\begin{document}
\title{NEUTRINOS AND THE WEAK INTERACTIONS}
\author{B.G. Sidharth\\
International Institute for Applicable Mathematics \& Information Sciences\\
Hyderabad (India) \& Udine (Italy)\\
B.M. Birla Science Centre, Adarsh Nagar, Hyderabad - 500 063 (India)}
\date{}
\maketitle
\vspace{2mm}
\begin{flushleft}
\end{flushleft}
\vspace{3mm}
\begin{abstract}
We show that neutrinos and electrons share the same theoretical structure, and satisfy parallel relations particularly of the Large Number kind. We then argue that the neutrino can be described as a ``cold'' electron in a sense that is detailed, and thereby the weak interactions are indeed a weak form of electromagnetism.
\end{abstract}
\section{Introduction}
Some years ago the author's work successfully predicted a dark energy driven accelerating universe with a small cosmological constant, as also a small mass for the neutrino, about a hundred millionth that of the electron \cite{ijmpa,ijtp,qp,ffp2,csf,cu}. These results were subsequently confirmed by the observations of Perlmutter and co-workers and other teams \cite{perl,kirsh}, while the Superkamiokande experiments also confirmed the mass of the neutrino, with the predicted value. The cosmological model referred to uses fluctuations in particle numbers and deduced moreover otherwise empirically well known supposedly mysterious, inexplicable coincidences-- the so called Large Number Relations and the Weinberg formula \cite{weinberg}. The mass of the neutrino was based on its anomalous behavior (Cf. also \cite{jsp}), and was also deduced, again on the basis of fluctuations. In this scheme, given $N$ particles, $\sqrt{N}$ particles condense out of the background dark energy or Quantum Vaccuum as a result of fluctuations. In this theoretical model we deduce that the gravitational constant is given by
\begin{equation}
G = \frac{c^3l^2}{\hbar \sqrt{N}}\label{e1}
\end{equation}
In the sequel $l, m$ denote the Compton length and mass of a typical elementary particle like the pion. (It may be noted that in the Large Number sense, the distinction between, for example, the pion and electron or proton masses gets blurred.) 
More recently (\ref{e1}) was shown to be related to the universal underpinning of Planck scale oscillators \cite{fpl}. Incidentally, this relation explains the longstanding puzzling empirical Weinberg formula referred to above. For, using the so called Eddington formula
\begin{equation}
R = \sqrt{N} l\label{e2} 
\end{equation}
$R$ being the radius of the universe, which was deduced in the above cosmological scheme, in (\ref{e1}) and further remembering that the Hubble constant $H$ is given by $\frac{c}{R}$ we can immediately deduce 
\begin{equation}
m = \left(\frac{H \hbar^2}{Gc}\right)^{\frac{1}{3}}\label{e3}
\end{equation}
which is the Weinberg formula. It may be mentioned that the puzzling feature of (\ref{e3}) has been that a so called microphysical parameter viz., $m$ depends on a cosmological parameter viz., $H$. However it turns out, as can be seen from (\ref{e1}), that $G$ itself has a distributional cosmological character so that, ultimately both the left side and the right side of (\ref{e3}) are purely microphysical (Cf.ref.\cite{fpl} for details).
\section{Neutrinos and the Weak Interaction}
With the above background, we now investigate the neutrino and weak interactions. We start by following Hayakawa \cite{hayakawa} to balance the gravitational force and the Fermi energy of the ``cold'' background neutrinos and further identify it with the intrinsic energy of the neutrinos to get
\begin{equation}
\frac{GN_\nu m^2_\nu}{R} = \frac{N_\nu^{2/3}\hbar^2}{m_\nu R^2} = m_\nu c^2\label{e4}
\end{equation}
where $m_\nu$ is the neutrino mass. From (\ref{e4}) we can immediately deduce that
\begin{equation}
m_\nu = 10^{-8} m_e , N_\nu \sim 10^{90}\label{e5}
\end{equation}
Both the relations in (\ref{e5}) are known to be correct.\\
We then use the fact that due to the fluctuation in the number of nutrinos, we have an energy which is the inertial energy again:
\begin{equation}
\frac{\bar g^2 \sqrt{N_\nu}}{R} \approx m_\nu c^2\label{e6}
\end{equation}
where $\bar g^2$ gives the weak interaction coupling constant.\\
Interestingly there is a similar relation for the electrons (Cf.ref.\cite{hayakawa})
\begin{equation}
\frac{e^2 \sqrt{N}}{R} = mc^2\label{e7}
\end{equation} 
From (\ref{e6}) and (\ref{e7}) on using (\ref{e5}) we get
\begin{equation}
\bar g^2 / e^2 \sim 10^{-13}\label{e8}
\end{equation}
which ofcourse is again known to be correct.\\
We have thus recovered from theory the well known values of the weak coupling constant and the neutrino mass. We would next like to show that there is a complete parallel between the Large Number Relations for elementary particles with similar relations for the neutrino. We start with the simplest relation, which can be easily verified
$$N_\nu m_\nu = Nm = M = 10^{55}gm,$$
$M$ being the Mass of the universe.
We next return to the fact used above in (\ref{e4}) and consider the equality of the gravitational mass of a particle due to the remaining $n$ particles with the inertial mass of the particle
\begin{equation}
\frac{Gnm^2}{r} = mc^2\label{e9}
\end{equation}
In (\ref{e9}), if $n$ is replaced by $N$ and $r$ is replaced by the radius of the universe, we get the mass of an elementary particle like the pion. On the other hand if in (\ref{e9}) we replace $n$ by the number of neutrinos $N_\nu$ instead of $N$ then we recover the mass of the neutrino. Finally if we take $n = 1$ and $r = l_P$, the Planck scale we recover the Planck mass $m_P$, which indeed is to be expected because as Rosen had shown, the Planck mass black hole is a universe in itself \cite{rosen}.\\
Similarly we see the complete parallel between (\ref{e6}) and (\ref{e7}). To proceed further we consider (\ref{e1}) in an alternative form viz.,
\begin{equation}
\hbar = \frac{Gm^2\sqrt{N}}{c}\label{e10}
\end{equation}
For the neutrino number and neutrino mass given in (\ref{e5}), (\ref{e10}) gives
\begin{equation}
\hbar' = \frac{Gm^2_\nu \sqrt{N_\nu}}{c} = 10^{-12} \hbar\label{e11}
\end{equation}
(\ref{e11}) shows that the magnetic moment of the neutrino is given by
\begin{equation}
\mu_\nu \sim 10^{-11} \, \mbox{Bohr \, magnetons}\label{e12}
\end{equation}
Indeed (\ref{e12}) is consistent with observation \cite{kolb}. That is for the neutrino we have effectively $\hbar'$ given by (\ref{e11}), instead of $\hbar$. It is then simple to verify that the analogue of the Eddington formula (\ref{e2}) applies for the neutrinos viz.,
$$R = \sqrt{N_\nu} l_\nu,$$
where $l_\nu = \frac{\hbar'}{m_\nu c}$, the neutrino analogue of the Compton length.\\
It has been shown on the basis of black hole radiation life times that we have
\begin{equation}
\frac{Gm^2}{l} = \frac{\hbar}{T} , T = 10^{17}sec\label{e13}
\end{equation}
where $T$ is the life time of the universe (Cf. also \cite{sivaram}). Indeed (\ref{e13}) is just a variant of the Weinberg formula, and can now be interpreted as the fact that the gravitational self energy of the elementary particle, viz., $\frac{Gm^2}{l}$ has a life time of the order of the age of the universe, due to the Uncertainty Principle. It can immediately be verified that for the neutrino we have the equation
\begin{equation}
\frac{Gm^2_\nu}{l_\nu} = \frac{\hbar'}{T}\label{e14}
\end{equation}
In the author's model, it has been shown that the pion can be considered to be an electron positron bound state so that we have
\begin{equation}
l = \frac{e^2}{m_e c^2}\label{e15}
\end{equation}
where $l$ is the pion Compton wavelength. Similarly one could consider the pion to also be the bound state of a quark anti-quark in QCD so that we have
\begin{equation}
\frac{g^2}{m_q c^2} = l\label{e16}
\end{equation}
where $m_q$ is the quark mass and $g^2$ is the strong interaction coupling constant. There is an immediate analogue of (\ref{e15}) and (\ref{e16}) for the neutrino viz.,
\begin{equation}
l_\nu = \frac{\bar g^2}{m_\nu c^2}\label{e17}
\end{equation}
Finally it may be pointed out that there is an immediate analogue of the Weinberg formula (\ref{e3}) viz.,
\begin{equation}
m_\nu = \left(\frac{H \hbar'^2}{Gc}\right)^{1/3}\label{e18}
\end{equation}
It must be mentioned that these analogues like (\ref{e6}), (\ref{e11}), (\ref{e14}), (\ref{e17}) and (\ref{e18}) between the neutrino and an elementary particle are not mere numerical coincidences. This is because the various relations for the elementary particles are the result of a theoretical structure, and are not mere accidents. What the foregoing means is that the neutrino has a similar theoretical structure.\\
To see this in greater detail, we note that in the case of the Planck scale underpinning for the universe of elementary particles, as was discussed in an earlier communication \cite{psu}, we have,
$$r = \sqrt{N \Delta x^2}$$
\begin{equation}
kl^2_P \equiv k \Delta x^2 = \frac{1}{2} k_B T\label{e19}
\end{equation}
where $k_B$ is the Boltzmann constant, $T$ the temperature, $r$ the extent and $k$ which resembles the spring constant is given by
$$\omega^2_0 = \frac{k}{m}$$
where $\omega_0$ is the frequency of a Planck mass viz.,
$$\frac{m_P c^2}{\hbar}$$
In the case of elementary particles it was shown that with $r \sim l$ the pion Compton wavelength we get from (\ref{e2})
\begin{equation}
k_BT = \frac{m^3c^4l^2}{\hbar^2} = mc^2,\label{e20}
\end{equation}
This as noted agrees with the Hagedorn temperature for elementary particles. For the neutrino a similar argument using the above equations including (\ref{e19}) gives, with the neutrino parameters $m_\nu, l_\nu$ and $\hbar'$ substituted in (\ref{e20}),
\begin{equation}
k_BT = m_\nu c^2\label{e21}
\end{equation}
Equation (\ref{e21}) gives for the neutrino mass
\begin{equation}
T \sim 1^\circ K\label{ex}
\end{equation}
which corresponds to the ``cold'' cosmic background temperature. This is completely consistent with our starting point in (\ref{e4}), where we consider the Fermi energy of the ``cold'' cosmic neutrinos. Infact the Fermi energy term in (\ref{e4}) (or the temperature (\ref{ex})) is the only difference between elementary particles and neutrinos - this is what leads to different values for $m_\nu, N_\nu$ etc. as compared to $m, N$ etc.
\section{Discussion}
We have seen that the weak interactions given by the coupling constant in (\ref{e8}) is a parallel of the electromagnetic interaction. Ofcourse in the standard electroweak theory \cite{taylor} the neutrino mass is taken to be zero. However after the Superkamiokande experiments, it has been realised that some modification in the standard model is required. We have seen above that it is the ``cold'' cosmic background or equivalently the Fermi energy which gives the neutrino its mass on the one hand and the weak interaction on the other. Further as can be seen from (\ref{e6}) and (\ref{e7}) the origin of the weak and the electromagnetic interaction is the same viz., the fluctuation in particle number in the universe.\\
Indeed based on these fluctuations we could even get the gravitational interaction, that is (\ref{e1}). Infact equating the fluctuational energy of electromagnetism (\ref{e7}) to the gravitational energy given in (\ref{e9}) with $N$ replacing $n$ and $R$ replacing $r$, we have
$$\frac{Gm^2 \sqrt{N}}{R} = \frac{e^2}{R}$$
that is
\begin{equation}
\frac{Gm^2}{e^2} \approx \frac{1}{\sqrt{N}}\label{e22}
\end{equation}
Equation (\ref{e22}) is the well known relation giving the ratio of the gravitational and electromagnetic coupling constants.\\
If we observe the parallel in the equations (\ref{e15}), (\ref{e16}) and (\ref{e17}), we can interpret (\ref{e17}) as describing a bound state of two neutrinos. The result is a particle of Compton wavelength $l_\nu$, that is a heavy particle of mass $10^4 m$. Such a particle would ofcourse be very shortlived. Indeed particles of this order of mass, for example the $*\gamma$ resonances are known \cite{hag}.\\
Finally it must be noted that the much smaller mass of the neutrino - approximately a vanishing mass - causes the four component Dirac electron equation to split into two component neutrino equations, as in standard theory, and thus gives the neutrino its handedness.\\
In summary we have shown that the neutrino can be described as a ``cold'' (old) electron.

\end{document}